# On the Enhanced Reverse Beta Processes in Graphene-Iron Composite Nanostructures at High Temperatures in Strong Magnetic Field

## Reginald B. Little

**Abstract**
Strong dense many-spin interactions have been proposed to organize novel orbital dynamics (the Little Effect) for novel chemical and catalytic phenomena. The recent determinations of the relativistic and quantum Hall effects of carriers in graphene under strong magnetic confinement have substantiated the Little Effect. Moreover such nonclassical phenomena under the stronger magnetic confinement of ferro-nanocatalysts are here shown to organize reverse beta processes and possibly pycnonuclear reactions under high temperature and high-pressure conditions. Such processes have implications for reverse beta reactions and nuclear reactions within the earth's interior and new technologies for carbon nanotube-ferrometal and nanographene-ferrometal composites.

**Introduction**
In 2000, a physical basis for ferromagnetic carbon was determined on the basis of defects, disorder, distortion, strain and pentagon-heptagon ring defects in graphene during its catalyzed nucleation and growth into carbon nanotubes (CNT) by ferro-nanocatalysts [1]. The ferromagnetism of the forming graphene and CNT was shown to assist the CNT coupling to the ferrocatalyst for the concerted organization of many carbon atomic orbitals into molecular graphitic orbitals by polarized spin currents and spin waves and their consequent dynamical electromagnetic (the Little Effect) induction between the catalyst and nucleating and growing CNT. The resulting dynamical magnetic fields on the molecular –nanometer length scales of the catalyst at the synthesis temperature (700-900 $^{o}$C) were shown to organize and orchestrate the CNT formation under the coulombic driving force of electron transfer between the catalyst and carbon atoms. These dynamical magnetic and electric fields of the nanocatalyst due to quantized spin currents and spin waves were determined to effect quantum Hall type interactions (even at the synthesis temperature) between the forming graphene and the underlying ferrocatalyst that provide the needed quanta of energy and momenta to accelerate and organize the ring currents in the forming graphene and CNT. In support of this, other investigators have recently experimentally and computationally demonstrated quantum Hall type effects in graphene even at room temperature [2,3]. The determined level spacing between Landau levels extrapolates such quantum effects to temperatures as high as 2800 K [3,4]. Moreover and in addition to the quantum dynamics, many investigators have experimentally and computationally demonstrated the relativistic nature of the quasiparticle in graphene for the Dirac spin character of the electrons and holes [4,5,6,7]. The electron velocities in graphene have been determined to exceed $10^{-2}$ c (1% the speed of light) [5,6,7]. Such relativistic speeds are here suggested to be even greater for nanographene due to the dangling bonds, surface energy and surface tension. Here it is demonstrated that the energies are even greater during the bond rearrangement for forming graphene and CNT due to the energetic accumulation from many broken bonds (even > 100,000 fractured chemical bonds) by antisymmetry in the strong magnetism.

Strong perpendicular external magnetic fields (20-45 teslas) have been shown to effectively confine the Dirac relativistic spins in graphene [8,9]. Even stronger transient ferromagnetic (hundreds of teslas) confinement by exchange and correlation with underlying ferrometals are here suggested to allow even tighter confinement of much higher energy Dirac spins. These recent findings provide strong support of the strong, dense spin-induced orbital dynamics known as the Little Effect and its account of CNT, diamond and reverse beta processes [10].

These recent nonclassical revelations of graphene have led to a remarkable union of condense matter physics and high-energy physics. On the basis of such relativistic quantum electrodynamics in graphene, here it was discovered that the ferrocatalyst provides very strong magnetic fields of several hundred Teslas during the catalytic growth of graphene for extremely confining these relativistic quasiparticles. The broken bonds, nanosize and curvature contribute to even greater energies of the electronic motion. Unlike molecular lasing with photon coupling many excited states for stimulating emission, in these magnetized condensed materials many (millions) excited spins by antisymmetry can be coupled through the very strong exchange and correlation. The resulting coupled relativistic quasiparticles can accumulate and concentrate (or dissipate) the energy of millions of these particles into a few particles with strong magnetic confinement by the surrounding ferromagnetic lattice. Because the graphene and the ferrocatalysts are excellent absorbers of hydrogen, such dynamic magnetic confinement of high energy carriers and absorbed $p^+$ in graphene-iron nanocomposites under synthesis conditions are here determined to provide a beautiful environment for reverse beta processes, infrequent nuclear reactions and other transmutation reactions. Because the interior of the earth is hot, magnetic and contains iron, and trace carbon and hydrogen, such reverse beta processes by enhanced electroweak effects in the iron – carbon composite are predicted in the mantle-core of the earth. Many future technologies and applications may develop as a result of such enhanced reverse beta in the ferrometal-graphene nanocomposites.

The internal processes within the interior of the earth's core and mantle have been debated by many scientists for many years. Some scientists have hypothesized concerning the existence and origin of possible heat sources within the earth's interior [12]. The actual compositions of the earth's outer and inner cores have been pondered [11,13]. Various hypotheses concerning the terrestrial core composition have been derived and argued. Various theories have been put forth. The hypotheses concerning the composition of the earth's core have been tested with the various seismological measurements through its center [14] and HPHT simulations of core conditions by studies using laser heating within diamond anvil cell [15,16]. The composition and properties of the earth's core have also been computationally studied. Such analyses of the earth are consistent with its core being mostly iron. On the basis of this iron composition, the magnetic properties of the earth's interior have also contributed to various theories of its composition and interior heating [17]. Here a new theory for internal thermal processes and dynamics within the earth is given, a new model for testing the theory is given and new data are presented. Here a new method of external magnetization is put forth for providing valuable analysis of possible internal processes within the earth's mantle and core on the basis of the Little Effect.

Some have reasoned that the earth is not internally heated but simply hot from the time of its formation by the insulation due to the surrounding mantle silicates [19]. Other theories of internal heating by chemical origins have been suggested on the basis of heats of solution and dissolution under gravitational driven buoyancy dynamics within the molten outer core that is sandwiched between the rotating solid inner core and rotating solid mantle [20]. On the basis of these various studies it is thought that the earth's core is mostly iron with various possible impurities. Some investigators have proposed slow terrestrial fusion as a possible source of heating within the earth's interior [21,22]. Other investigators have reasoned that radioactivity decay of some heavy elements contributes to heating the earth's interior. Recently, the enhanced reverse beta processes and neutron activated elemental transmutations were discovered [10] during magnetized, high temperature, high current and conductive activation of some metal hydrides by many body correlated spin interactions (the Little Effect). Here it is discovered and demonstrated that ferromagnetic metal-graphene hydrides nanocomposites also provide a natural strong magnetically correlated environment for such discovered accelerated reverse beta processes of internal electrons and protons and the consequent neutron catalyzed elemental transmutations. Furthermore, it is suggested that such processes may occur within the hot and compressed hydrogenous, carbonaceous iron in the mantle-core of the earth for both terrestrial fusion and fission nuclear reactions. Hydrogen has been shown to form stable phases with hcp iron at high pressures with dramatic alteration of the magnetism of the iron [18].

In this paper, results are presented from such experiments on magnetized hot hydrogenated graphene-iron composites that test for the occurrence and enhancement of reverse beta processes in the composite. This previously proposed model was tested using a similar system that has been used to show diamond nucleation and growth (also an internal terrestrial process) from catalyzed carbon black at atmospheric pressure and high temperatures (920 $^\circ$C) with Fe catalyst and a hydrogenous atmosphere in strong static magnetic field of 17 Tesla [23]. Such low energy nuclear reactions were explored by strongly magnetizing Fe, carbon black, hydrogen gas at 920 $^\circ$C.

**Procedure**
A mixture of iron and carbon black was placed inside a sealed quartz vessel and installed within the bore of a DC magnet. The vessel was purged with flowing Ar and heated to 920 $^\circ$C. Hydrogen was then flowed over the Fe and carbon black and the Ar flow was terminated. After 10 minutes of reduction of the catalyst by flowing hydrogen, the DC magnet was slowly ramped up to 17 Tesla. $CH_4$ was also introduced with the $H_2$ flow over the Fe and carbon black. The system was maintained at 920 $^\circ$C and 17 T in the flowing $H_2$ and $CH_4$ for 2.5 hours. Afterward the Ar flow was restarted and the $CH_4$ and $H_2$ flows were terminated. The magnetic field was ramped to 0 tesla. The sample was then immediately removed from the magnet and cooled to room temperature by applying water to the outside of the quartz tube. The Fe-C samples were analyzed by secondary ion mass spectroscopy (SIMS). It is important to note that here strong magnetization of iron at high temperatures is introduced as a system to mimic some of the internal conditions in some parts of the earth, although with slower kinetics due to the lower pressure.

**Results and Discussions**

The high temperature (920 °C) annealing of the C, Fe sample under the strong magnetic field and the methane-hydrogenous atmosphere causes catalytic chemical transformation of the carbon as the Fe dissolves the carbon and reprecipitates it as some graphite and some diamond. The magnetization shifts the bonding probability toward more diamond [23]. The magnetization also allows the accumulation of high energy by antisymmetry of broken bonds in the defective distorted graphene on the iron catalyst for substantial electroweak and electromagnetic interactions between the high-energy relativistic electrons and absorbed protons in the graphene-iron composite for reverse beta processes and neutron formation. The neutrons are absorbed by the surrounding hydrogen and carbon with the formations of $^2$D, $^3$T, and $^{13}$C. SIMS analysis determined the presence of these nuclides in anomalous relative amounts.

SIMS is used to study the Fe, C, H matrix before and after the high temperature magnetic annealing in order to observe possible changes in isotopes and nuclide intensities. A typical SIMS of the C, Fe, H matrix before (control) the high temperature magnetization in hydrogenous atmosphere is shown in Figure 1. The SIMS of the C, Fe, H matrix after (sample) the high temperature, magnetic annealing in hydrogenous atmosphere is shown in Figure 2. Various differences develop in the mass spectrum of the sample after its magnetic high temperature treatment in hydrogenous atmosphere. In general, on the basis of the SIMS, distinct isotopic differences exist between the sample and control for hydrogen and carbon, which are the major components of the samples. Furthermore, the mass intensities (for many nuclides lighter than iron) are greater after the high temperature magnetization. On the one hand, the mass intensities (for some nuclides heavier than iron) are less after the high temperature magnetization. Moreover, it is important to note the existence of unchanged mass intensities of some nuclides with masses between 32-54 Daltons in order to demonstrate the precision of the analysis and furthermore to demonstrate the transformations of certain suitable nuclides and the intact ablation effects of some nuclides, while other nuclides remained unaltered in their natural relative abundances. The observed changes in relative amounts of some nuclides before and after are evidence of low energy nuclear reactions during the magnetization of the hot iron-graphene structures in the hydrogenous atmospheres.

The greater mass intensity of 2 Da in the sample relative to the control is consistent with $^2$D formation in the sample after its high temperature magnetic treatment in hydrogenous atmosphere. The $^2$D intensity in the sample is $6 \times 10^2$ and the intensity in the control is about $2.5 \times 10^2$. It is important to note the almost equal mass intensities of $^1$H (1 Da) in both sample and control. The sample is also richer in mass 3 Da. No mass 3 Da is observed in the control. The mass 3 Da may be either $^3$T or $^3$He. The mass 1 Da intensities are about $2.3 \times 10^6$ and $2.7 \times 10^6$ in the sample and control, respectively. Gas phase clustering is not as significant due to the use of Cs$^+$ primary beam during the SIMS. The C, H, Fe matrix has lower carbon clustering in the gas phase upon Cs$^+$ bombardment relative to an $O_2^+$ primary ion bombardment. However, the $Cs^+$ primary ions can knock clusters from the films during the SIMS. But very little clustering occurs in the gas phase after the bombardment during SIMS. The observed ablated intact clusters from the matrix include C-H (13 Da), $C_2$ (24 Da), $C_3$ (36 Da), $C_4$ (48 Da), C-O (28 Da), O-H (17 Da), $C_5$ (60 Da), $C_6$ (72 Da), $C_7$ (84 Da), and $C_8$ (96 Da). If $H_2$ formed in the gas phase during the SIMS, then its intensity would be as great as the mass intensities for $C-H_2$ (14 Da) (7 X $10^4$), $C_2H_3$ (27 Da) (2 X $10^4$), $C_3H_3$ (38 Da) (3X$10^3$), C (12 Da) (4 X $10^6$) and H (1 Da) (2.3 X $10^6$) masses. But the observed intensity from possible gaseous clustering to cause mass 2 Da is 6 X $10^2$, much less than such possible gas phase clustering of $^{12}C$ and H to $^{12}$C-H (13 Da) (2.6 X $10^7$). It is not likely that the carbon atoms in the matrix are bound to two hydrogen atoms. On this basis, either H atoms are not clustering in the gas phase during SIMS and C and H atoms are clustering or not much clustering of either H atoms or H and C atoms occurs in the gas phase during the SIMS. Here the latter is suggested on the basis that the observed 13 Da is due to ablated $^{12}$C-H and the observed 2 Da peak is due to $^2$D ablated from the matrix. The $^2$D formed by reverse beta and neutron absorption processes to form $^2$D during the high temperature magnetic treatment of the sample. Mass 2 Da has to be $^2$D because not much $H_2$ exist in the matrix, but the $^2$D can exist in the lattice of the sample due to nuclear reactions during the magnetic high temperature treatment. So the observed CH, $C_2H_2$, and $C_3H_2$ result mostly from the ablation of intact molecules from the matrix. Where as intact $H_2$ cannot be ablated from the matrix and not much H atoms cluster to $H_2$ in the gas phase during the SIMS, $^2$D can be ablated, explaining the 2 Da mass peak in the sample. These mass intensities of carbon clusters may also be due to CD, $C_2D$, and $C_3D$, which are further consistent with the $^2$D formation and the reverse beta processes during the magnetic annealing of the sample. For example, the ratios of $C_3/C_3D$ in the sample and control are about 2 and 4, respectively. These differences in $C_3/C_3D$ between the sample and control cannot be explain by gas phase clustering differences of $C_3$ with 2H atoms or even the ablation of more intact $C_3H_2$ from the sample matrix relative to the control matrix. The $^2$D in the sample is however slightly over twice (2) the amount of $^2$D in the control, consistent with the observed relative ratio of $C_3/C_3D$ in the control to sample.

Based on the C-H mass (13 Da) intensity relative to the H mass (1 Da) intensity, the $CH_2$ (14 Da) intensity is 10-100 times smaller than the C-H (13 Da) mass intensity in the sample. The $CH_2$ form by possible C-H clustering with H in the gas phase during the SIMS. Therefore, if the H clustered in the gas phase, then the $H_2$ would be 10-100 times smaller than the H, for a required mass 2 Da signal on the order of 6 X $10^4$ in the sample. But the actual observed of mass intensity of 2 Da in the sample is much smaller, having an intensity of 6 X$10^2$. Therefore gas phase clustering cannot explain the observed 2 Da mass intensity relative to other ablated clusters $CH_2$, $CH_3$, $C_2H_2$, and $C_3H_2$, where some gas phase clustering is possible. The mass 2 Da is therefore due to $^2$D and not H atoms clustering in the gas phase during SIMS to $H_2$. It should be pointed out that this mass 14 Da may be N although the system was purged of $N_2$ before the annealing process and the mass 14 Da exist in both sample and control with equal intensity. Furthermore, if H atoms clustered in the gas phase for the sample during SIMS then it should also cluster in the gas phase for the control because there is the same amount of gas phase H in both sample and control, 2.3 X $10^6$ and 2.7 X $10^6$, respectively. But much less mass 2 Da is observed in the control relative to the sample. It is not possible that the sample and control can have the same H atom mass intensity in the gas phase yet the H atoms in the sample cluster more in the gas phase than the gas phase clustering of H atoms in the control during the SIMS. Therefore the observed greater mass 2 Da in the sample is due to $^2$D and not $H_2$. For the same reasons (with even greater power) given here for mass 2 Da peak, the mass 3 Da is due to $^3$T and/or $^3$He and not $H_3$ clusters. Perhaps the strongest evidence that the masses of 2 Da and 3 Da in the sample are not due to gas phase clustering is that these masses are not seen in the SIMS of the control matrix, in spite of the control having as much C and H mass intensities as their observed intensities in the sample. The observed 2 Da and 3 Da masses in the sample are a result of the neutron formation by the reverse beta process during the high temperature magnetic treatment, and the absorption of these neutrons by $^1$H $\rightarrow$ $^2$D and $^2$D $\rightarrow$ $^3$T during the hot, magnetic annealing process.

The mass intensity of $^{12}$C (12 Da) is as high as the H (1 Da). This is consistent with the matrix being mostly carbon and hydrogenated carbon. The mass 12 intensities are different in the sample (4.2 X $10^6$) and control (1.5 X $10^7$). This drop in $^{12}$C in the sample may be due to its transmutation to $^{13}$C. The carbonaceous nature of the matrix is consistent with the observations of various carbon molecules $C_2$, $C_3$, $C_4$, $C_5$, $C_6$, $C_7$, and $C_8$ in both sample and control matrixes. It appears that these molecules are knocked away intact from the matrix and do not form by gas phase clustering during the SIMS. The mass intensities for these carbon clusters are seen in both the sample and the control with about equal intensities relative to the mass 12 Da of $^{12}$C. It is important to note the very high intensity of mass 13 Da in the sample. Relative to mass 12 Da, mass 13 Da may be mostly $^{12}$C-H, which is knock away intact from the matrix during SIMS. Unlike similarities in $C_2$ (24 Da), $C_3$ (36 Da), $C_4$ (48 Da), $C_5$ (60 Da), $C_6$ (72 Da), $C_7$ (84 Da), and $C_8$ (96 Da) mass intensities relative to mass 12 Da for the samples and the control, it is quite remarkable that the observed mass intensity of 13 Da relative to the intensity of mass 12 Da differs between the sample and the control. The sample has slightly higher observed ratio of mass 13 Da / mass 12 Da intensity relative to the ratio of mass 13 Da /mass 12 Da ratio observed in the control. The higher observed mass 13 Da intensity in the SIMS of the sample is a result of the greater $^{13}$C in the sample relative to $^{13}$C in the control. Clustering effects cannot explain the mass 13 Da difference between the sample and the control, because gas phase clustering should occur from $^{12}$C and H knocked loose from both sample and control matrices with equal probability, especially considering the sample and control reveal similar concentrations of $^{12}$C and $^1$H. So why should the gas phase clustering in the sample be greater? Better yet both sample and control were heated in hydrogen, only the sample was strongly externally magnetized. Why should the sample release more intact $^{12}$C-H? This extra mass 13 Da and its nonclustering origin are consistent with the observed similar $^{12}$C-$^{12}$C-H (25 Da) relative to $^{12}$C-$^{12}$C (24 Da) mass differences between the sample and the control. If the mass 13 Da difference was due to different intact ablations of $^{12}$C-H clusters then similar differences in intact ablation of $^{12}$C-$^{12}$C-H should exist between the sample and the control. But the $^{12}$C-$^{12}$C-H peak does not differ between the sample and the control. So the 13 Da peak is not all due to cluster ions from the sample; some of the 13 Da peak is due to excess $^{13}$C in the sample. The really conclusive evidence of more $^2$D in the sample relative to the control is found by further comparing the mass 26 Da ($C_2$D) intensities between the sample and control. Whereas no difference exists for mass 25 Da between the sample and the control due to similar intact ablation of $C_2H$ between the sample and the control, the mass 26 Da ($C_2D$) intensity in the sample is much greater than its mass intensity in the control, reflecting the higher levels of formed $^2D$ in the sample and a resulting higher amount of $C_2D$ ablated from the sample relative to the control. On the basis of the reverse beta processes and neutron absorption by the $^{12}C$ and $^1H$, more $^{13}C$ and $^2D$ exist in the sample relative to the control. Such higher levels of $^{13}C$ by $^{12}C$ transmutation is consistent with the observed high levels of $^{13}C$-$^{13}C$ to $^{12}C$-$^{12}C$ and $^{12}C$-$^{13}C$ to $^{12}C$-$^{12}C$ in the sample relative to the control as evidenced by different relative mass signals of 25 Da and 26 Da relative to the mass intensity of 24 Da in the sample relative to the control. These results are only possible if some of the H in the sample has been converted to neutrons, which absorb into the surrounding atomic nuclei to form new nuclides of greater mass by 1 Da. Therefore more masses of 2 Da, 13 Da, 25 Da, and 26 Da are observed in the sample relative to the control due to more $^2D$, $^{13}C$, $^{13}C$-$^{12}C$ and $^{13}C$-$^{13}C$ in the sample relative to the control.

The anomalous levels of $^2D$, $^3T$ and $^{13}C$ in the sample after its high temperature magnetic annealing is a result of enhanced reverse beta and other nuclear processes of hydrogen absorbed within the iron-graphene nanocomposites. The external magnetization of the iron-graphene system causes transient ferromagnetism under the high temperature conditions. The graphene is coupled to the iron for its strong magnetization to several hundred teslas due to its coupling with the iron lattice. The graphene under these conditions exhibits quantum Hall effects under the strong magnetic field of the underlying iron catalyst. The magnetism via exchange with the iron is much greater in the graphene of this composite structure relative to recently explored by external magnetization of graphene (45 teslas) with an electromagnet [3]. Therefore the quantum Hall effects in the more strongly magnetic graphene-iron composite here extends to much higher temperatures than the observed room temperature quantum Hall effects under 45 Teslas by external magnetic field on graphene using an electromagnet [3]. This strong magnetization of the graphene by the attached iron and the broken bonds (Fe-Fe, Fe-C, C-C, C-H) under the high temperature conditions result in the greater relativistic character of the electrons, protons and holes with even greater, tighter cyclotron motion and confinement under the stronger magnetization of the iron catalyst. These greater relativistic electrons and protons under the conditions in the graphene-iron composite are able to interact via electroweak effects to form neutrons. Here it is important to consider the validity of these low energy nuclear reactions on an energetic basis in such possible ferrocatalyst-graphene systems. Consider for example a 6 nm Fe particles (about 6500 iron atoms) and its ability to catalyze the growth of a CNT at 750 °C at a rate of over 200 nm/s under the growth conditions. Such an iron particle must handle more than 100,000 carbon atoms per sec to form the CNT at this remarkably high growth rate. In other words, the 6 nm Fe nanoparticle (6500 atoms) has activated per second over 100,000 C-C bond, which are some of the strongest possible chemical bonds. All of the energy in these 100,000 very strong C-C bonds amounts to more than the energy to activate a nuclear reaction between protons, deuterons and electrons even under the strong force and columbic $p^+ - p^+$ repulsion. The discovery here is that the strong magnetic field of the iron allows simultaneous activation of millions of chemical bonds by antisymmetry and the strong exchange and correlation of the electrons of these many broken bonds can accumulate and concentrate the energy into a few particles with the magnetic confinement of the resulting relativistic particles. The discovery here is that the transient magnetic order by the imposed external magnetic field on the ferromagnetic catalysts has some low probability to organize, confine, accumulate (by antisymmetry) and concentrate this energy via Dirac spin many body interactions for orchestrating such for nuclear reactions within the nanocatalyst-graphene composite. These discovered phenomena can lead to new energy technologies for ferrometal filled CNT and ferrometal-graphene composites. This discovery can also lead to new transmutation technologies for handling radioactive waste. This prospect is consistent with the recent demonstration by other investigators of the alteration of the nuclide decay lifetime by a metal lattice [33]. This discovery can also explain nuclear transformation within the interior of the earth and other planets.

Here the roles of such reverse beta processes and neutron generation for activation of elemental transmutation and radioactive decay within the earth are put forth to supplement the older radioactive decay fission theories and fusion theories for internal terrestrial heating. The data here support the theory of terrestrial nuclear reactions as a heat source within the earth. The proposed iron content of the earth's core with the tremendous pressures and temperatures and carbon and hydrogen impurities within the core provides a suitable environment for such magnetocatalytic reverse beta processes within the interior mantle-core of the earth. The large hydrogen absorptions of graphene and iron and high temperature and high-pressure effects on hydrogen absorption [18] are supportive of such reverse beta processes in these hydrogenated ferrometal-graphene systems and the hydrogenated mantle-core of the earth.

In 2004, an electroweak model was proposed for the reverse beta processes in activated metal hydrides and extended this effect for such reverse beta processes in the earth's interior for generating terrestrial nuclear activity [10]. In this model, the possible concentration of solar neutrinos [24] through the earth's core by the earth's magnetic field contributes an actively important ingredient to such geo-nuclear reactions. Moreover in this model, the strong magnetic field persists in the core and mantle materials even at the high interior temperatures due to the offsetting high pressure conditions within the deep interior [15]. The high magnetic field, high temperature and high pressure were postulated to catalytically enhance reverse beta processes involving dissolved hydrogen and solar neutrinos for generating neutrons, which combine with hydrogen to generate deuterium, tritium, which form $^3$He and $^4$He [10]. Such neutrons further activate the transmutations of other surrounding elements. The high temperature, high pressure conditions drive chemical bond rearrangement and melting of the iron media. The high temperature and high pressure sustain the magnetic order of the core materials for ferromagnetism in the iron in spite of the temperature being higher than the Curie temperature [33]. The magnetism and consequent antisymmetry of the broken bonds result in the accumulation of chemical energy and the organization of the energy to organize novel chemistry in the mantle like diamond, and other gem formation and to organize the reverse beta process of electrons and protons for neutron formation. The resulting neutrons are absorbed by the surrounding elements for elemental transmutations and the continuous redistribution of energy by lighter elements forming heavier elements about Fe and heavier elements breaking (fission) into iron on the basis of the thermodynamic stability of iron nucleus.

The analogy here between the diamond formation and nuclear reactions is important since both are shown to occur inside the earth. It is important here to compare the analogous magnetic effects on diamond, graphene and CNT syntheses with the here discovered magnetic effect on these reverse beta phenomena. Similar effects of magnetic force on hot iron catalyst for the activation and breakage of iron, carbon and hydrogen bonds, for the antisymmetric slowing of lower order hybrid ($sp^2$) chemical bond formation, for the radical accumulation and higher order collisions, for higher order rehybridization ($sp^3$) to form diamond by the Little Effect lead to higher order many spin interactions to rehybridizing electronic hydrogenous orbitals into nuclear motions and symmetries forming neutrons to contribute to enhanced beta processes but at higher energies. Similar antisymmetry and energy accumulations and collective motions lead to higher order multi-polarized $e^-$, $p^+$, neutrino interactions for enhanced reverse beta processes to form neutrons (as with diamond formation) are demonstrated here [10]. Furthermore similar conditions (high temperatures, high pressures, catalytic ferrometal and strong magnetization) inside the earth cause such magnetocatalysis for both diamond formation and reverse beta processes. It is important to note further that the strong magnetic field and its effects on these processes have also been reported by others via magnetic monopoles from intense laser fields for both igniting nuclear reactions and organizing the diamond formation [25]. The strong magnetic exchange between the ferromagnetic catalyst and the $e^-$, $p^+$ and neutrinos prevent $\gamma$ exchange between the $e^-$ and $p^+$ for hydrogen formation, thereby allowing greater probability of the $e^-$ collapse onto the $p^+$, preventing reverse beta. Strong magnetic exchange and antisymmetry allow many body interactions of the spins for accumulating and concentrating chunks of potential energy for the nonclassical acceleration of a few spins to extremely high velocities for such processes of diamond and neutron formations but under lower mechanical pressures. This nano-graphene iron composite also exhibits dense plasmons and relativistic electron and hole velocities for such fruitful environments for diamond formation and nuclear reactions.

Such many body phenomena for the relativistic nature of the spins in graphene have recently been observed and determined to support this prior theory of Little. The resulting quantum electrodynamics and electromagnetism of the high spin systems allow diamond to form. Likewise, the resulting quantum electrodynamics between the hydrogen and the lattice allow electroweak interactions in the high spin system for reverse beta processes. At the atmospheric pressure conditions, the strong external magnetic field enhances the transient magnetic order in the ferrometal in spite of the high temperatures, thereby allowing these electromagnetic and electroweak processes within the earth but at lower frequencies, the higher pressures sustain magnetic organization despite high temperatures [15]. The same phenomena (the formation of diamond [26], the formation of gems, the formation of high pressure phase crystals, the lowering of melting point [25] and here the reverse beta processes) resulting from pressure induced magnetism within the high temperature interior of the earth can occur at lower pressures even atmospheric pressure (although at slower rates) if an external magnetic field is provided at the lower pressure for transient magnetic order in spite of the high temperature. Hence novel phenomena have been observed by magnetizing media above the Curie temperature for the Little Effect and frustration of the Hedvall Effect [10,23]. By magnetizing the media at high temperature, the Hedvall Effect is frustrated and chemical-catalytic conditions are created akin to the earth's core.

In order for spin currents and spin waves of the catalysts to affect graphene by the consequent dynamical magnetic field, the graphene itself should be magnetic. Moreover in 2000, ferromagnetism was proposed in the carbon nanostructures as well as the metal [1]. Such ferromagnetic carbon can be realized on the basis of defects, disorder, and distorted C bonds on the nanoscale. Such magnetism in carbon leads to novel magnetocatalytic phenomena. Furthermore, recently researchers have demonstrated ferromagnetism in carbon under such defective, disorder and distorted nanocarbon [27-31]. Moreover, recently many researchers have demonstrated the highly organized energetic effects in the ferrometal-graphene by their observations and computations of Dirac nature of carriers in graphene [5-7] and strong magnetic field, the relativistic nature of the carriers [7,8], the strong confinement of the Dirac fermions under strong perpendicular magnetic fields [9] and the quantum Hall effect in graphene sustained to 2880 $^{o}$C in magnetic fields greater than 20 Teslas. Compressed Fe was also seen to exhibit relativistic phenomena [32]. In these systems, the surface tension and energy in addition to the attached carbon-graphene causes pressure on nano-iron to cause relativistic description of is carriers. Here it is demonstrated that the influence of quantum Hall effects of the nanocatalysts on the formation of graphene into CNT and also the influence of relativistic effects in the electronics and spin currents and waves between the ferrocatalyst and forming graphene/CNT on the growth mechanism of the CNT. Such quantum Hall effects in graphene under the huge magnetic exchange with the contact with the ferromagnetic metal catalyst (several hundred teslas) are here justified in promoting novel nuclear reactions in ferrometal-graphene composites under the high temperature and external strong magnetization. Here it is demonstrated that such relativistic Dirac fermions in broken, disordered, and distorted surface graphene under strong magnetization of several hundred tesla by ferromagnetic exchange with underlying ferrometal provides beautiful conditions for reverse beta processes of adsorbed hydrogen due to enhanced electroweak interactions by the relativistic motion of fermions ($e^-$, $p^+$, neutrinos) strong magnetic confinement, antisymmetric accumulation of fermions and high accumulated concentrated energy for correlated collective motion for organized reverse beta processes.

**Conclusion**
On the basis of the SIMS analysis of the Fe,C matrix before (control) and after (sample) high temperature (920 $^{o}$C) annealing in hydrogenous atmospheres at 17 tesla, unusual higher levels of masses 2 Da, 3 Da and 13 Da are observed in the sample relative to the control. On the basis of the similar H and C levels in the sample and control matrixes, these mass differences cannot be all due to clustering effects during the SIMS.

These higher levels of masses 2Da, 3Da, and 13 Da are attributed to isotopes $^2$D, $^3$T, $^3$He and $^{13}$C due to neutron adsorption processes of H and C in the matrix by neutron generated by reverse beta processes during the high temperature annealing under strong magnetization. These reverse beta processes during this high temperature magnetic annealing in hydrogenous atmospheres result from chemical energy stored in the broken bonds under antisymmetry during the carbon, metal and H bond rearrangements as Fe adsorb carbon, $CH_4$ and $H_2$ and forms grapheme and some diamond. These reverse beta phenomena in Fe-graphene, Fe-CNT systems may have future technological applications as energy sources based on electroweak effects and the energy liberated during reverse beta processes. Such low energy nuclear reactions may explain the magnetism, internal heating and terrestrial nuclear reactions in the interior of the earth on the basis of such reverse beta processes in the iron media.

**Acknowledgements**

With External Gratitude to GOD!

In Loving Memory of Lewis Parks.

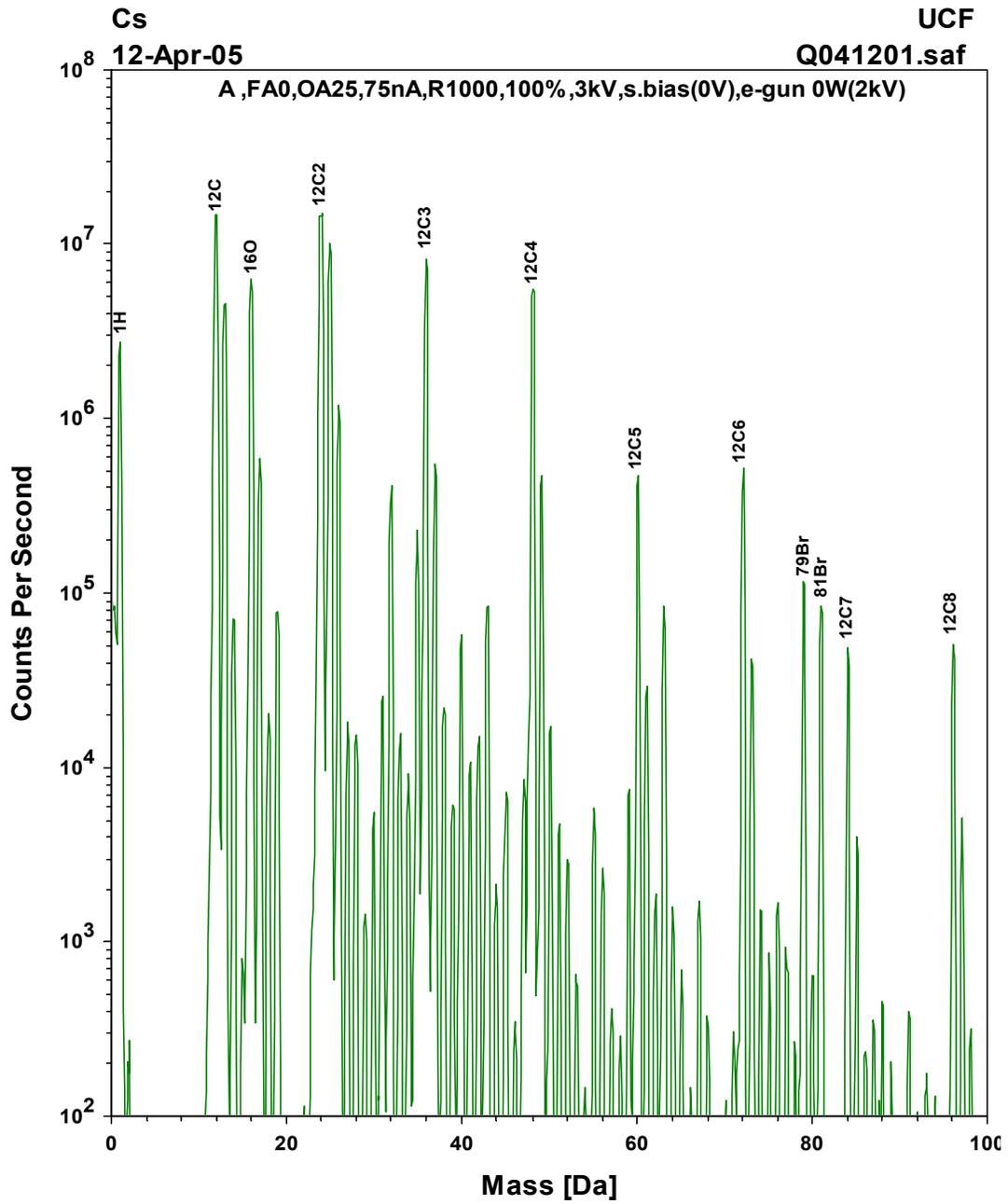

**Figure 1**

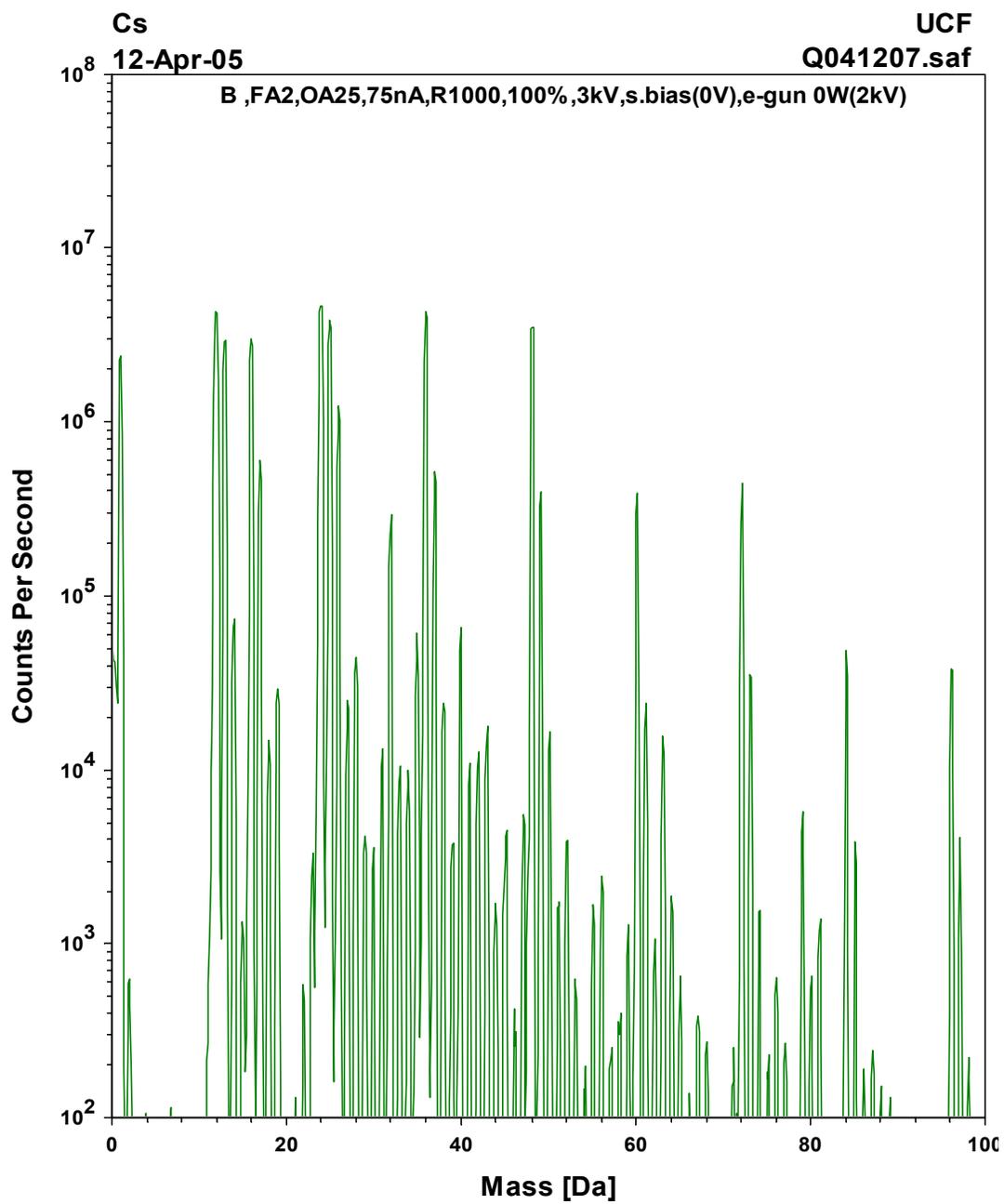

**Figure 2**